\documentclass[
    ,final            
  ]
  {aipproc}

\layoutstyle{6x9}

\usepackage{latexsym}

\newcommand{\be}{\begin{equation}}
\newcommand{\ee}{\end{equation}}
\newcommand{\bea}{\begin{eqnarray}}
\newcommand{\eea}{\end{eqnarray}}
\newcommand{\bref}[1]{(\ref{#1})}
\newcommand{\LGB}{\mathcal{L}_\mathrm{GB}}
\newcommand{\dz}{\partial_z}
\newcommand{\Mpl}{M_\mathrm{Pl}}
\newcommand{\Mph}{M_{\Phi}}

\begin{document}

\title{Gauss-Bonnet Brane World Gravity with a Scalar Field}

\author{Stephen C. Davis}{
  address={Institut de Physique Th\'eorique, \'Ecole Polytechnique
F\'ed\'erale de Lausanne, CH--1015 Lausanne, Switzerland}
}

\begin{abstract}
The effective four-dimensional, linearised gravity of a brane world
model with one extra dimension and a single brane is
analysed~\cite{GBgrav,GBgrav2}. The model includes higher order
curvature terms (such as the Gauss-Bonnet term) and a conformally
coupled scalar field. Large and small distance gravitational laws are
derived. In contrast to the corresponding Einstein gravity models, it
is possible to obtain solutions with localised gravity which are
compatible with observations.  Solutions with non-standard large
distance Newtonian potentials are also described.
\end{abstract}

\maketitle


\section{Second Order Gravity and Brane Worlds}

In the Randall-Sundrum II (RS) brane world scenario~\cite{RSII}, we live on
3+1 dimensional brane embedded in a 4+1 dimensional bulk spacetime. As
a result of the warping of the fifth dimension, the effective
gravitational theory on the brane closely resembles that which is
observed in our universe (except at very small distances). In this
paper we will investigate an extended version of this scenario, which
includes a conformally coupled bulk scalar field, $\phi$, and an
action which is of quadratic order in curvature. We will
consider $Z_2$-symmetric solutions of the form 
\be
ds^2= e^{2A(z)}(dx^2_4 +dz^2) \ \ \mathrm{with} \ \ 
A=-\ln(1+|z|/\ell) \ \ \mathrm{and} \ \ 
\phi/A=u = \mathrm{constant}
\ee
 which is the simplest
generalisation of the RS model. The brane is located at $z=0$. We will
be mainly interested in solutions with $\ell >0$ (which is required to
localise gravity) and we take  $u\ell >0$ in order to avoid bulk singularities.
We find that when the conformally coupled scalar field is added to the
RS model, solutions with localised gravity can no longer be
found. However, this problem can be fixed by including higher order
gravity terms~\cite{selftune}, as we will discuss in later sections.

In four dimensions, the gravitational field equations (for the vacuum)
are taken to be $G_{ab} + \Lambda g_{ab} = 0$. These can be derived by
looking for a rank 2 curvature tensor which (i) is symmetric, (ii) is
divergence free, and (iii) depends only on the metric and its first
two derivatives. In five
dimensions the above conditions are satisfied by
$G_{ab} + 2\alpha H_{ab} + \Lambda g_{ab} = 0$, where $H_{ab}$ is the
second order Lovelock tensor~\cite{Lovelock}. $H_{ab}$ can be obtained
from the variation of an action containing the Gauss-Bonnet term
\be
\LGB = R^{abcd} R_{abcd} - 4 R_{ab}R^{ab} + R^2 \ .
\ee
Energy momentum is conserved in the corresponding gravitational
theory and its vacuum is ghost-free (just as in Einstein gravity).
Note that $H_{ab}$ is the only quadratic curvature term
which satisfies the above three conditions. In four dimensions its
contribution to the field equations is trivial, and so it is usually
ignored.

It is natural to expect the Gauss-Bonnet term to appear in the action
of any five-dimensional theory, since even if it is not part of the
fundamental theory, it is likely to be generated by quantum gravity
corrections. A further reason for including the term in our brane
model is that brane worlds are motivated by string
theory, and the Gauss-Bonnet term also appears in low energy effective
string actions.

Our model also includes a scalar field. In a string theory context
this would correspond to the dilaton, or a moduli field coming from the
compactification of other extra dimensions. As with the curvature
terms, it is natural to include higher order
scalar kinetic terms in the action. We
will consider the general second order contribution 
\be
\mathcal{L}_2 = 
c_1 \LGB - 16 c_2 G_{ab}\nabla^a\phi \nabla^b \phi 
+ 16 c_3  (\nabla\phi)^2 \nabla^2 \phi - 16 c_4 (\nabla\phi)^4  \ .
\label{L2}
\ee

If $\phi$ is the dilaton, not all of the above coefficients are fixed
by string theory. In fact low energy string theory actions suggest
that $R^2$, $R_{ab} R^{ab}$, $R (\nabla\phi)^2$ and  $R \nabla^2 \phi$ are
also possible.
However we will not consider these since they all give ghosts at high energy
(but they are not actually ruled out since we are dealing with a low energy
effective action).
If $\phi$ is a moduli field, the coefficients can be determined from the shape
of the compactified dimensions.

We could also include third and higher order scalar kinetic terms,
although for simplicity we will not consider them. In
this case the full bulk action (in the string/Jordan frame) is
\be
S_\mathrm{Bulk} = \frac{1}{2\kappa^2} \int d^5x  \sqrt{-g}  
e^{-2\phi}\left\{ R - 4\omega (\nabla \phi)^2
+ \mathcal{L}_2  - 2\Lambda \right\} \ .
\ee
$\Lambda$ is the bulk cosmological constant.

The brane can be treated as a boundary of the bulk spacetime. In order
for the Einstein-Hilbert action to consistent, we need to add the
Gibbons-Hawking term~\cite{GibbHawk} to the 
boundary action. Similarly, we will need a suitable boundary
contribution to go with the Gauss-Bonnet term~\cite{Myers}
\be
\LGB^{(b)} 
= \frac{4}{3}( 3K K_{ac}K^{ac} - 2 K_{ac}K^{cb}K^a{}_b -K^3)
- 8 G^{(4)}_{ab}K^{ab} \ .
\ee

Including the second order scalar field terms as well, the brane
contribution to the action is
\be
S_\mathrm{brane} = - \frac{1}{\kappa^2}\int d^4x \sqrt{-h} e^{-2\phi} \, 
\left\{ 2K + \mathcal{L}_2^{(b)} + T \right\}
\ee
where $T$ is the brane tension and
\be
\mathcal{L}_2^{(b)} = c_1 \LGB^{(b)} 
- 16 c_2 (K_{ab}- K h_{ab})D^a\phi D^b \phi
- 16 c_3 (n \! \cdot \! \nabla \phi) 
\left(\frac{1}{3}(n \! \cdot \! \nabla \phi)^2 + (D\phi)^2\right) \ .
\ee
Variation of the action gives the generalised Israel junction 
conditions for the brane~\cite{GBgrav,BCGB}. These do not depend on the brane
thickness (this is not true for other second order gravity terms).

\section{Linearised Brane Gravity}

In general, a perturbation of the bulk metric will alter the position
of the brane, and it is then necessary to change coordinates to put
the brane back at $z=0$. We avoid this problem by working in a gauge
in which the brane remains at $z=0$. This is achieved by considering a
general perturbation of the bulk metric, and then using the components
of the bulk field equations which are normal to the brane to determine
$g_{\mu z}$ and $g_{zz}$. We obtain the perturbed metric~\cite{GBgrav2}
\be
ds^2 = e^{2A}\left[
(\eta_{\mu \nu} + \gamma_{\mu \nu}) dx^\mu dx^\nu + dz^2 \right] 
- \ell e^A dz (dx^\mu \partial_\mu + dz\dz) 
\left( N_1(u) \psi + \frac{2}{u}\varphi \right) \ ,
\label{pertmet}
\ee
with
\be
\gamma_{\mu \nu} = 
\bar \gamma_{\mu \nu} + 2(\zeta+\psi) \eta_{\mu \nu} 
-2N_2(u) \frac{\partial_\mu \partial_\nu}{\Box_4} \psi \ ,
\label{gdef}
\ee
where $\partial^\mu \bar \gamma_{\mu \nu} = 0$ and $\eta^{\mu \nu}\bar
\gamma_{\mu \nu} = 0$. The perturbed scalar field is $\phi =
u A + \varphi$. The fields $\psi \propto u \gamma-8\varphi$
and $\zeta$ are linear combinations of $\varphi$ and 
$\gamma=\eta^{\mu \nu} \gamma_{\mu \nu}$. The values of the various
coefficients and the definitions of $\psi$ and $\zeta$ can be found in
ref.~\cite{GBgrav2}.

The bulk field equations give the wave equation for the transverse
traceless part of the four-dimensional metric 
\be
\mu_\gamma(u) \left(\dz^2 - (3 -2u) \ell^{-1} e^A \dz 
+ f_\gamma^2(u) \Box_4\right)\bar \gamma_{\mu \nu} = 0 \ .
\label{BgGB}
\ee
Note that there are no third or fourth order derivatives in the
equation, despite the fact we have field equations which are of quadratic
order in the curvature. The corresponding equation for the RS model
has $u=0$ and $\mu_\gamma=f^2_\gamma=1$.

If either of $\mu_\gamma$ or $f^2_\gamma$ are negative, the kinetic
term for $\bar \gamma_{\mu \nu}$ will have the wrong sign in the
effective action, so the bulk will have graviton ghosts~\cite{us}. This
requirement restricts the allowable ranges of the model's parameters.

The effective gravitational law on the brane is obtained from the junction
conditions. For $\bar \gamma_{\mu\nu}$ we find
\be
2\mu_\gamma \dz \bar \gamma_{\mu \nu}+m_\gamma^2  \Box_4 \bar \gamma_{\mu \nu}
= -2\kappa^2  \left\{S_{\mu\nu} 
- \frac{1}{3}\left(\eta_{\mu\nu} 
- \frac{\partial_\mu \partial_\nu}{\Box_4}\right)S \right\}
\label{bcgGB}
\ee
where $m^2_\gamma = 8c_1(1-2u)/\ell$ and $S_{\mu\nu}$ is the
perturbation of the brane energy momentum tensor. For the RS
model we would just have
$\dz \bar \gamma_{\mu \nu} = - \kappa^2  \{S_{\mu\nu}- \cdots \}$.  

If $m^2_\gamma<0$ then either the effective Planck mass on the brane
is negative, or the vacuum has a non-trivial solution with
spacelike momenta, i.e.\ the graviton spectrum includes a tachyon and
the solution is unstable.

The graviton wave
equation~\bref{BgGB} is solved (for spacelike momenta) by
\be
\bar \gamma_{\mu \nu} \propto 
e^{-ip \cdot x} e^{-(2-u)A} K_{2-u}\left(f_\gamma p \ell e^{-A}\right) \ .
\label{gsol}
\ee
Using a small $p$ series expansion, we see (for $u<1$) that
\be
\dz \bar \gamma_{\mu\nu} \approx 
-\frac{\ell \, f_\gamma^2 \, p^2 }{2(1-u)} \bar\gamma_{\mu\nu}
\label{bigr}
\ee
for large distance scales ($1/p \gg \ell f_\gamma$). So we will have 
$\Box_4 \bar\gamma_{\mu\nu} \propto -\{ S_{\mu\nu} - \cdots \}$, which
is similar to the RS scenario, and will give a $1/r$ contribution to
the Newton potential. The extra $\Box_4 \bar\gamma_{\mu\nu}$ term in the
junction conditions gives a similar contribution at short distances
(unlike the RS model), so the inclusion of  higher
gravity terms weakens the short distance gravity constraints on the model.

The behaviour of the scalar mode $\psi$ is qualitatively similar. Its
bulk field equation is the same as eq.~\bref{BgGB}, but with
different parameters $\mu_\psi$ and
$f^2_\psi$. Its junction condition is
$2\mu_\psi \dz \psi + m_\psi^2 \Box_4 \psi = -\kappa^2 S$.
As with the graviton modes, ghosts and tachyons are present for some
parameter ranges.

The remaining degree of freedom, $\zeta$, is pure gauge in the
bulk, but its behaviour on the brane is given by
$m_\zeta^2 \Box_4 \zeta = - \kappa^2 S$. It can be interpreted as the
brane-bending mode.

\section{Effective Four Dimensional Gravity}

Putting all the junction conditions together, we obtain the following
expression for the induced metric perturbation
\be
\gamma_{\mu \nu}(p) = 2 \kappa^2\Biggl(
 D_\gamma(p) \left\{S_{\mu\nu} 
- \frac{1}{3}\eta_{\mu\nu} S \right\}
+ D_\psi(p) \eta_{\mu\nu} S
+\eta_{\mu\nu} \frac{S}{m_\zeta^2 p^2}
\Biggr)
\label{prop}
\ee
where
\be
D_i(p) = \left\{m_i^2 p^2 + 2\mu_i f_i p 
\frac{K_{u-1}(f_i \ell p)}{K_{2-u}(f_i \ell p)} \right\}^{-1} \ .
\label{propD}
\ee
We have omitted the $p_\mu p_\nu/p^2$ dependent terms for simplicity.
The three contributions to eq.~\bref{prop} correspond to the graviton
modes, the bulk scalar, and the `brane-bending' mode. 
The Newton potential and the effective four-dimensional graviton
propagator can be extracted from the above expression~\bref{prop}.

If $u<1$, then the leading order behaviour of the
perturbation~\bref{prop} at large distances ($1/p \gg \ell f_{\gamma,\psi}$) is
\be
\gamma_{\mu \nu}(p) \approx
\frac{2}{\Mpl^2 p^2} \left\{S_{\mu\nu} - \frac{1}{2}\eta_{\mu\nu} S \right\}
+ \frac{1}{\Mph^2 p^2}\ \eta_{\mu\nu} S \ .
\label{propBD}
\ee
This describes Brans-Dicke gravity, with graviton mass $\Mpl = \tilde
m_\gamma/\kappa$  and scalar mass $\Mph$ given by
\be
\frac{1}{\Mph^2} = 2\kappa^2 \left(\frac{1}{6\tilde m^2_\gamma} 
+ \frac{1}{\tilde m_\psi^2} + \frac{1}{m_\zeta^2}\right) \ ,
\label{Ms}
\ee
where $\tilde m^2_i = \mu_i f_i^2/(1-u) + m^2_i$. To
avoid conflict with solar system measurements we need $\Mph \gg
\Mpl$~\cite{damour}. This can be achieved for suitable parameter
choices. If $m^2_\gamma$ is non-zero we
also obtain Brans-Dicke gravity at shorter distances, but
with different values of $\Mpl$ and $\Mph$.

As an example we will take
$\omega=-1$ and $c_i=\alpha$ which would correspond to $\phi$ being
the dilaton (with some extra symmetries).

For Einstein gravity ($\alpha=0$), there is just one solution, with
$u=\infty$. This is similar to a negative warp factor ($\ell <0$)
solution, and does not give localised gravity. Furthermore, when
higher order terms are included, $m^2_\gamma$ is negative, and so the
solution develops a tachyon. So we see that in this case the inclusion
of higher gravity terms has not solved the solution's problems. In
fact it has made them worse.

When the second order gravity terms are turned on ($\alpha > 0$),
two extra solutions appear
\be
\frac{\phi'}{A'} = u 
= \frac{3}{2} \pm \sqrt{\frac{3}{4} + \frac{\ell^2}{8\alpha}} \ .
\ee
The upper sign choice always gives a solution with instabilities,
while the solution with the lower choice is stable if $u < 1/2$. We will just
consider the latter solution for the remainder of this section.

Although the behaviour of the bulk graviton and scalar modes is
qualitatively similar, the higher gravity terms give different
contributions to the coefficients in the two wave equations. The 
degeneracy between scalar and graviton modes is broken by higher order gravity.
In particular we can have
\be
f_\gamma^2 = \frac{1-2u}{1-u} \ll f_\psi^2 = \frac{3}{3-2u}
\ee
for the above solution if $u \approx 1/2$.

When $1/p \gg \ell f_\psi$ (large distances), we find the couplings
for the effective four-dimensional gravity are
\be
\Mpl^2 = 8 M^3 \alpha \ell^{-1} (2-u) f_\gamma^2
\ee
\be
\Mph^2 = 8 M^3 \alpha \ell^{-1} (3-2u)
\ee
It is therefore possible to obtain $\Mph \gg \Mpl$ if the solution is
fine-tuned to have $f_\gamma \ll 1$. This allows solar system
constraints (from linearised gravity) to be satisfied. This is despite
the fact that for the underlying five dimensional theory
$\Mpl^{(5)} \sim \Mph^{(5)}$. Note that the above fine-tuning of
parameters is in addition to usual brane world fine-tuning of
the cosmological constant, $\Lambda$, and the brane tension, $T$.

At intermediate ($\ell f_\gamma \ll 1/p \ll \ell f_\psi$)  and short
($1/p \ll \ell f_\gamma$) distance scales, we find 
$\Mph^2 \leq 3\Mpl^2$. However if these length scales are of geographical size,
there will be no problem, since short distance constraints on
scalar-tensor gravity are weak.

\section{Modified Large Distance Gravity}

So far we have assumed that the effects of the scalar field are
smaller than the warping of space
time (i.e\ $\phi'/A'=u<1$). However if this is not true, it is
possible to obtain solutions with non-standard Newton potentials at large
distances. 

If $1<u<2$ then the expression~\bref{bigr} is no longer valid, and instead
we have
\be
\dz \bar \gamma_{\mu\nu} \approx -\bar\gamma_{\mu\nu}
\frac{2 \, \Gamma(u-1)}{\ell \, \Gamma(2-u)}
\left(\frac{p\ell}{2}\right)^{4-2u} 
\label{bigr2}
\ee
when $1/p \gg \ell f_\gamma$. Substituting eq.~\bref{bigr2} into the junction
condition~\bref{bcgGB}, we find that $\bar \gamma_{\mu\nu}$ gives a
non-standard $1/r^{2u-1}$ contribution to the large distance Newton
potential. We now have only massive graviton bound states and no
localised zero mode. However
we can still obtain four-dimensional gravity at short distances
from the $\Box_4$ terms in the junction conditions.  The scalar modes have
similar behaviour.

The resulting large and short distance gravity has some resemblance to
the Dvali-Gabadadze-Porrati (DGP) model~\cite{DGP}, especially for the
special case of $u=3/2$. We then have $1/r^2$ contributions to the
large distance Newton law, as would normally occur in five-dimensional gravity.

Closer examination of the model reveals that it has more in common
with quasilocalised brane gravity models~\cite{quas} than the DGP
model. Unfortunately this type of model either has ghosts or an
unacceptably large scalar coupling~\cite{GRSprob}. By looking at the
expression for the effective scalar mass~\bref{Ms} we see that 
$\Mph \gg \Mpl$ is not possible unless $m_\zeta^2$ (or $m_\psi^2$) is
negative. Unfortunately if either of these parameters is negative the
theory will have a ghost (or a tachyon if $m_\psi^2<0$ and
$\mu_\psi >0$).

The RS model also has $m_\zeta^2<0$, but this is not a problem. This
model has two massless graviton zero modes, and an unphysical
graviscalar zero mode. The brane-bending ghost zero mode cancels the
graviscalar (a similar idea is used in the quantisation of QED).
However this is not possible in quasilocalised models since there
are no massless graviscalar or scalar states for the ghost to cancel
with. For our model it is possible to cancel the contribution of the massive
graviscalar modes by making $\psi$ a ghost~\cite{GBgrav2}, although in
contrast to the RS model, this is likely to be a physical ghost.


\begin{theacknowledgments}
I am grateful to the other conference organisers for their work in
producing such an enjoyable conference, and I thank
the Swiss Science Foundation for financial support.
\end{theacknowledgments}

\def\Journal#1#2#3#4{{#1} {\bf #2}, #3 (#4)}

\def\NPB{{\em Nucl. Phys.} B}
\def\PLB{{\em Phys. Lett.}  B}
\def\PRL{\em Phys. Rev. Lett.}
\def\PRD{{\em Phys. Rev.} D}

\end{document}